\documentclass[11pt]{article}
\usepackage{epsfig}
\newlength{\textbildsep}
\usepackage{amssymb}
\usepackage{amsmath}

\newtheorem{theorem}{Theorem} 

\newtheorem{corollary}{Corollary}

\newtheorem{fact}{Fact}

\input mssymb.tex

\def\R{\Bbb R}
\def\Z{\Bbb Z}

\begin{document}
\title{An explicit formula for the number of tunnels in digital objects}
\author{Valentin E. Brimkov\thanks{Mathematics Department, Buffalo State College, State University of New York,
1300 Elmwood Ave., Buffalo, NY 14222, USA.
Email: brimkove@buffalostate.edu.}
\ \ \
Angelo Maimone$^{\dagger}$ \ \ \ Giorgio Nordo\thanks{Dipartimento di Matematica, Universit\`a di Messina, 98166 Messina, Italy.
E-mail: $\{$angelo.maimone,giorgio.nordo$\}$@unime.it.}}

\date{}
\maketitle

\begin{abstract} 
An important concept in digital geometry for computer imagery is that of tunnel.
In this paper we obtain a formula for the number of tunnels 
as a function of the number of the object 
vertices, pixels, holes, connected components, and $2 \times 2$ grid squares. 
It can be used to test for tunnel-freedom  
a digital object, in particular a digital curve.

{\bf Keywords:} {\em  digital geometry, digital object, tunnel in a digital object}
\end{abstract}
 
\section{Introduction}
\label{intro}

An important concept in digital geometry for computer imagery is that of tunnel.
Intuitively, tunnels are locations in a digital object 
(that is any finite set of pixels/voxels in 2D/3D) through which
a `discrete path' can penetrate.
Tunnels play an important role in rendering 
pixelized/voxelized scenes by casting digital rays from the image to the scene
\cite{CohenKaufman}.
Thus it is useful to know if a digital object is tunnel-free or it has tunnels of certain type.
This is particularly interesting when dealing with digital curves or surfaces.
It may also be helpful to have an estimation for the number of tunnels (if any) in the 
considered object, possibly as a function of other object characteristics.
Such kind of information may help better understand the structure of the object.
With this in mind, our objective was to obtain a formula that relates basic parameters of 
a 2D digital object, such as the numbers of its pixels, vertices, holes, connected components, and tunnels. 

Results of this kind are of interest within several disciplines, such as digital geometry for computer imagery, 
image analysis, computer graphics, and combinatorial geometry.
Let us mention, for instance, the famous
Euler's formula $v-e+f=2$ that relates the number of vertices, edges, and facets of a polytope, 
and its applications to image analysis and digital geometry \cite{KLE2004}.
Other similar results are also available (see, e.g., Chapters 4 and 6 of \cite{KLE2004}).

The main result of this paper is the formula $t=v-2(p+c-h)+b$, where $t$ is the number of tunnels, 
$v$ the number of vertices, $p$ the number of pixels, $h$ the number of holes, 
$c$ the number of connected components, 
and $b$ the number of $2 \times 2$ grid squares (called 2-{\em blocks}) in a digital object. 
This equality implies corollaries for the important cases 
of simple closed digital curves, simple digital arcs, as well as for general digital curves.

In the next section we introduce some basic notions and notations.
In Section 3 we present our main results, and we conclude in Section 4.

\section{Preliminaries}
\label{prelim}

\subsection{Basic Definitions}
\label{sec:def}

In this section we recall some basic notions of digital geometry following
\cite{KLE2004}. The reader is also referred to \cite{KON2001,BRI2002,BBN}.

A regular orthogonal grid subdivides $\R^2$ into
unit squares called {\em pixels}, that are centered at the points of $\Z^2$.
Two pixels are called 0-{\em adjacent} if
they share a vertex, and 1-{\em adjacent} if they share an edge.
A digital object $S$ is a finite set of pixels.

In the following definitions $k=0$ or 1.
A $k$-{\em path} in $S$ is a sequence of pixels from
$S$ such that every two consecutive pixels are $k$-adjacent.
Two pixels of a digital object $S$ are {\em $m$-connected} (in $S$)
iff there is a $k$-path in $S$ between them.
A digital object $S$ is {\em $k$-connected} iff
there is a $k$-path connecting any two pixels of $S$.
The maximal (by inclusion) $k$-connected subsets of a digital object $S$ are called
(connected) $k$-{\em components} of $S$.
Clearly, components are nonempty and distinct components are disjoint 
with respect to $k$-adjacency.

\begin{figure}[t]
  \begin{center} 
\leavevmode
\epsfxsize=10cm
\epsffile{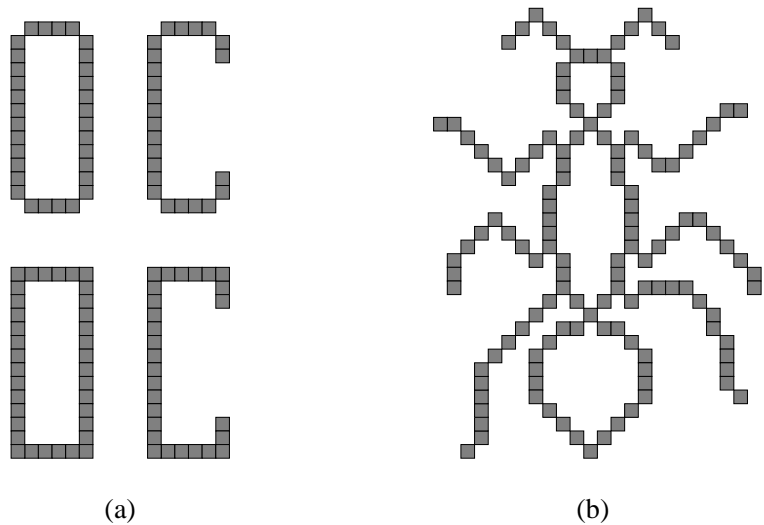}
\end{center} 
 \caption{a) {\em Top:} Digital curves with 0-tunnels. 
{\em Bottom:} Tunnel-free digital curves. Each `0' features a hole.  b) General digital curve with three holes.} 
\label{connect0}
\end{figure}

Let $M$ be a subset of a digital object $S$.
If $S \setminus M$ is not $k$-connected then
the set $M$ is said to be $k$-{\em separating} in $S$.
Now let $M$ be a finite digital object that is $k$-separating in $\Z^2$. 
The infinite 1-component of $\Z^2 \setminus M$ is called an {\em improper 1-hole} of $M$,
while the other (finite) 1-components of $S \setminus M$ are called {\em proper 1-holes} of $M$
(see Fig. \ref{connect0}).

Let a digital object $M$ be $1$-separating but not $0$-separating
in a digital object $S$. Then $M$ is said to have $0$-{\em tunnels}
(see Fig. \ref{connect0}a, top-left). 
For an object $M$ that is not separating in another object we can define a tunnel as 
a point that is a common vertex of two and only two pixwls of $M$ (see Fig. \ref{connect0}a, top-right).
A digital object without any $0$-tunnels is called
{\em tunnel-free} (Fig. \ref{connect0}a, bottom).

\subsection{Pixel Language}
\label{language}

To facilitate the further description, below we present a simple graphical `language'
(similar to Venn diagrams in set theory), which we call the {\em pixel language}.
We will consider different configurations of pixels,
as in each there will be a `key' pixel, highlighted in grey,
whose neighborhood is studied. 
Here are some key points of the pixel language 
(see for illustration figures \ref{connect1},\ref{connect2}, and \ref{connect3}).

\begin{itemize}
\item
Neighbors of the key pixel that {\em must} exist in the 
considered configuration will be drawn with normal continuous lines. 
\item
Pixels that {\em may} or {\em may not} exist in a configuration will be drawn with dashed lines.
Any subset of such pixels (in particular, no one or all of them) 
may belong to the configuration or may be missing. 
\item
Sometimes at least/at most/exactly one (or more) of these pixels will have to exist in a configuration. 
In order to keep our pixel language simple, we will prefer to add relevant explanations in the text rather than 
to introduce further special markings
(for other purposes it may be useful to do so).
\item
Grid positions that cannot contain pixels from the configuration will be marked by $\times$. 
\item
Sometimes we will assign labels to pixels
and with their help analyze certain possibilities.
\item
Existing path in a digital object connecting pixels from the configuration will be marked 
by a curve (as in some cases of figures 3 and 4).
\end{itemize}

\section{Tunnel formulas}
\subsection{Number of tunnels in arbitrary digital object}
\label{formula}
In this section we prove the following statement. 

\begin{theorem}
\label{Th1}
Consider a digital object $D \subset \Z^2$.
Let $p$ be the number of its pixels, $v$ the overal number of its pixels' vertices, 
 $h$ the number of its holes, $c$ the number of its connected components, 
$b$ the number of its 2-blocks, and $t$ the number of its $0$-tunnels. 
Then
\begin{equation}
\label{eq1}
t=v-2(p+c-h)+b.
\end{equation}
\end{theorem}
\noindent {\bf Proof} \
We use induction on the number of pixels. 
The statement is obviously true for a digital object consisting of a single pixel. 
We have $p=1$, $v=4$, $c=1$, and $h = b = t = 0$, which values satisfy formula (\ref{eq1}).
 
Assume that the statement is true for a digital object composed of $p$ pixels, where $p \geq 1$.
We will show that it is then true for any object composed by $p'=p+1$ pixels.
Consider such an object $D'$ and remove an arbitrary pixel $P$ from it. 
Then $D=D'-\{P\}$ is a digital object of $p$ pixels to which the induction hypothesis applies.
Let the number of its vertices, components, holes, 2-blocks, and tunnels be $v,c,h,b$, and $t$, respectively.
Then $t=v-2(p+c-h)+b$. We will see how adding pixel $P$ to $D$ can influence this last equality.
We aim to show that 
\begin{equation}
\label{eq2}
t'=v'-2(p'+c'-h')+b', 
\end{equation}
where $p'=p+1, v', c', h', b'$, and $t'$ are the counts 
of pixels, vertices, components, holes, 2-blocks, and tunnels of $D'$, respectively.
In doing so, we distinguish between 32 essentially different configurations
which we group into 10 cases, some of which involve 
subcases.\footnote{In fact, there are a number of configurations that are not explicitly elaborated here,
but all of them fall within the considered 10 basic cases and admit analogous characterization. Their description
is differed to the full-length paper. Note also that any configuration containing dashed pixel(s) and/or path(s),
actually represents a number of different configurations, one for each possibility,
as all of them feature analogous properties.}
We analyze all of them with the help of illustrations using our pixel language.
For the sake of better readability of the paper,
we display the illustrations within three figures (number 2, 3, and 4), as labels of subfigures match the 
numeration of the cases considered. Everywhere pixel $P$ is in dark grey.
We outline the main points of the proof. Details are diferred to the full length paper. 

Remember that throughout we have $p'=p+1$. 
Upon adding  $P$ to $D$, for the other parameters $v', c', h', b'$, and $t'$ 
we may have the following possibilities.

\medskip

{\bf Case 1} \ 
{\em $c$, $h$, and $b$ do not change.}

\begin{figure}[t]
  \begin{center} 
\leavevmode
\epsfxsize=8cm
\epsffile{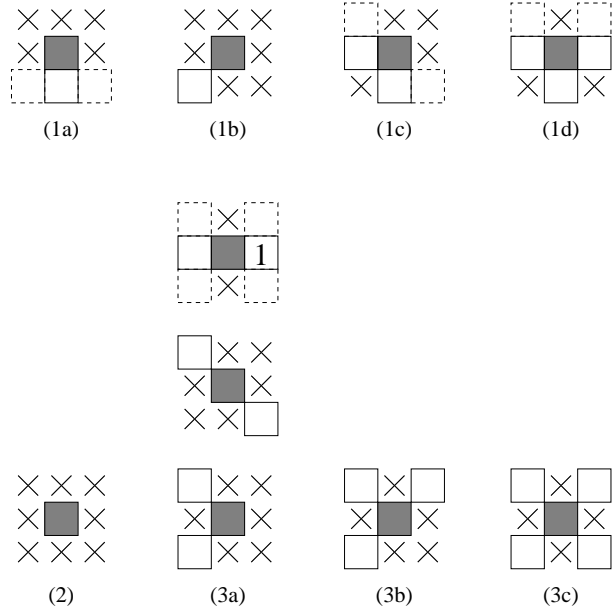}
\end{center} 
 \caption{Illustrations to the proof of Theorem \ref{Th1}. Cases 1, 2, and 3.} 
\label{connect1}
\end{figure}

The only possible configurations under these conditions are displayed in Fig. \ref{connect1}.
We have $c'=c$, $h'=h$, and $b'=b$.

In Case 1a we have $v'=v+2$ and $t'=t$.
In Case 1b, $v'=v+3$ and $t'=t+1$.
In Case 1c, $v'=v+1$ and $t'=t-1$.
In Case 1d, $v'=v$ and $t'=t-2$.

\medskip

{\bf Case 2} \ 
{\em $h$ and $b$ do not change, while $c$ increases by 1.}

Obviously, adding $P$ to $D$ can increase the number of components of $D'=D \cup \{ P \}$ 
only if $P$ is disjoint from $D$. 
Then $c$ will increase by 1, while $h$, $b$, as well as $t$ will not change
(see Fig. \ref{connect1} (2)). 

We have $v'=v+4$, $c'=c+1$, $h'=h$, $b'=b$, and $t'=t$. 

\medskip

{\bf Case 3} \ 
{\em $h$ and $b$ do not change, while $c$ decreases by 1,2, or 3.}

The possible configurations are displayed in Fig. \ref{connect1}.
We have $h'=h$ and $b'=b$.

In Case 3a we have $c'=c-1$, which is possible for two configurations displayed in Fig. \ref{connect1} (3a: top, middle, bottom). 
For the first configuration we have $v'=v$ and $t'=t$, while for the second and the third we have $v'=v+2$ and $t'=t+2$, 
respectively.  
In Case 3b we have $c'=c-2$. The possible configuration is the one in Fig. \ref{connect1} (3b).
We have $v'=v+1$, $t'=t+3$. 
In Case 3c we have $c'=c-3$ (Fig. \ref{connect1} (3c)).
We have $v'=v$, $t'=t+4$. 

\medskip

{\bf Case 4} \ 
{\em $b$ and $c$ do not change, while $h$ decreases by 1.}

\begin{figure}[t]
  \begin{center} 
\leavevmode
\epsfxsize=9cm
\epsffile{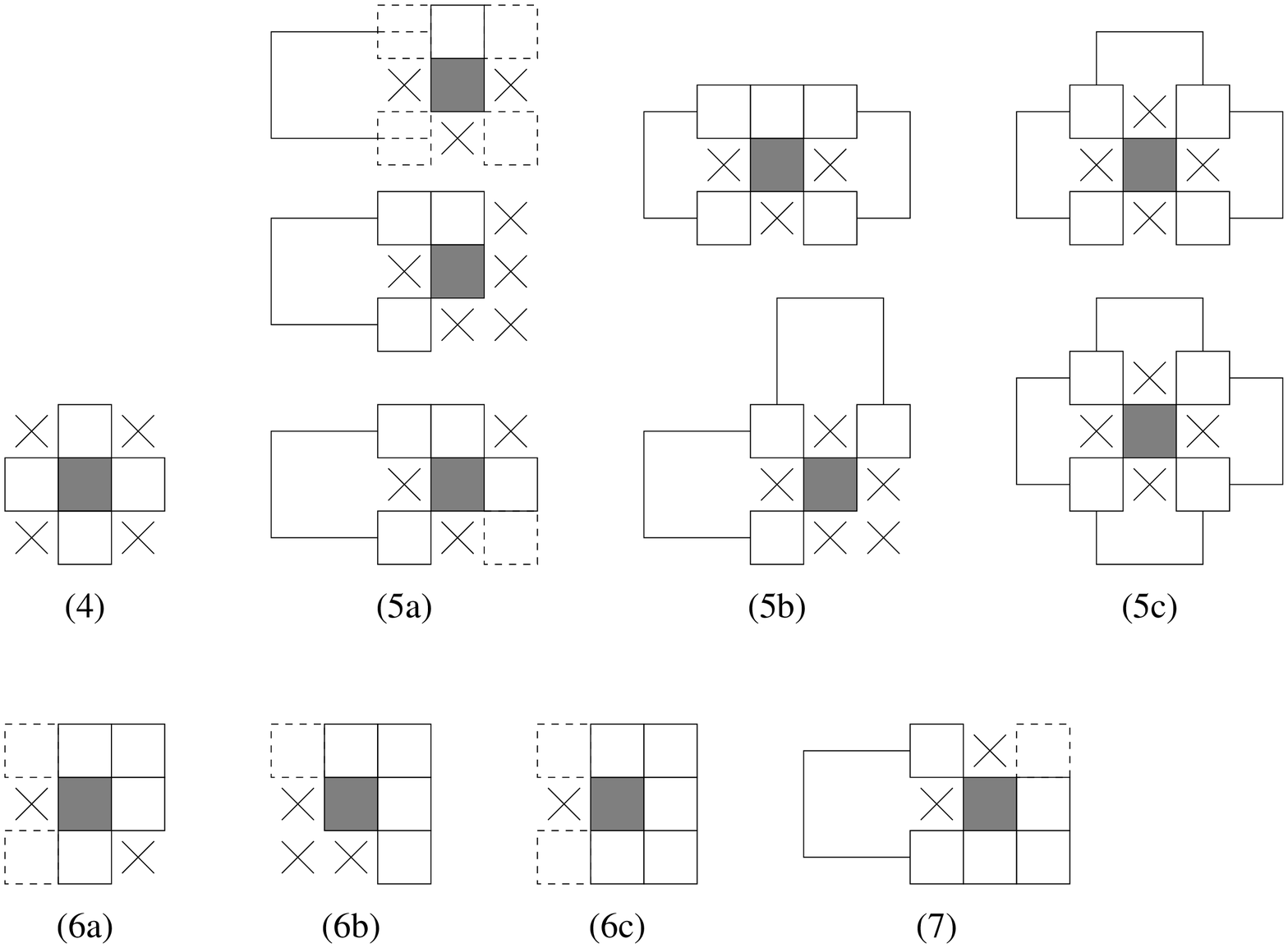}
\end{center} 
 \caption{Illustrations to the proof of Theorem \ref{Th1}. Cases 4, 5, 6, and 7.} 
\label{connect2}
\end{figure}

The only possible configuration is displayed in Fig. \ref{connect2} (4).
We have $h'=h+1$, $c'=c$, $b'=b$, $v'=v$, $t'=t-4$. 

\medskip

{\bf Case 5} \ 
{\em $c$ and $b$ do not change, while $h$ inecreases by 1,2,  or 3.}

The possible configurations are displayed in Fig. \ref{connect2}.
We have $c'=c$ and $b'=b$.

In Case 5a we have $h'=h+1$, for which there are three possible configurations displayed in Fig. \ref{connect2} (5a: top, middle, bottom). For the first configuration, $v'=v$ and $t'=t$. For the second, $v'=v+1$ and $t'=t+1$. For the third, $v'=v$ and $t'=t$ 
(note that adding $P$ creates a new tunnel but closes an existing one).

In Case 5b we have $h'=h+2$, which is possible for two configurations displayed in Fig. \ref{connect2} (5b: top, bottom). 
For the first one, $v'=v$ and $t'=t+2$, while for the second $v'=v+1$ and $t'=t+3$.  

In Case 5c we have $h'=h+3$, which may occur in two configurations (Fig. \ref{connect2} (5c)). 
For both we have $v'=v$  and $t'=t+4$.  
Note that in the first case three new holes may appear in a hole-free component, 
while in the second an existing hole is partitioned into four holes. 

\medskip

{\bf Case 6} \ 
{\em $c$ and $h$ do not change, while $b$ increases by 1 or 2.}

In the former case there are two possible configurations displayed in Fig. \ref{connect2} (6a,b). 
We have $c'=c$ and $h'=h$.
In Case 6a we have $b'=b+1$,  $v'=v$, and $t'=t-1$ (Fig. \ref{connect2} (6a)).
In Case 6b we have $b'=b+1$, $v'=v+1$, and $t'=t$ (Fig. \ref{connect2} (6b)). 
 
In the latter case, we have the configuration displayed in Fig. \ref{connect2} (6c). 
We have $b'=b+2$, $c'=c$, $h'=h$, $v'=v$, and $t'=t$. 

\medskip

{\bf Case 7} \ 
{\em $c$ does not change, while $h$ and $b$ increase by 1.}

The only possible configuration is displayed in Fig. \ref{connect2} (7).
We have $v'=v$, $h'=h+1$, $b'=b+1$, $c'=c$, and $t'=t+1$.

\medskip

{\bf Case 8} \ 
{\em $c$ does not change, $h$ decreases by 1, and $b$ increases by 1,2,3, or 4.}

\begin{figure}[t] 
  \begin{center} 
\leavevmode
\epsfxsize=9cm
\epsffile{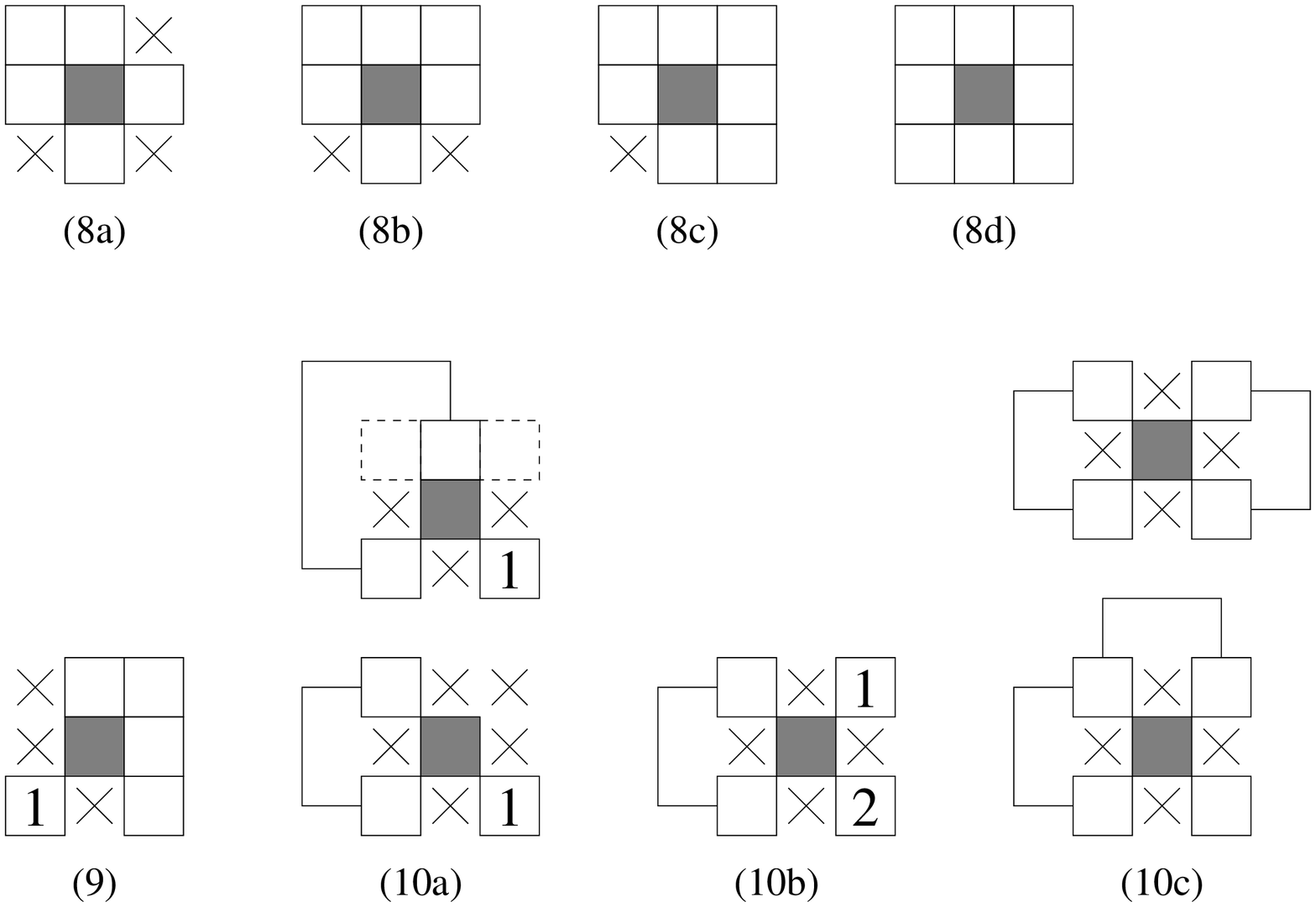}
\end{center} 
 \caption{Illustrations to the proof of Theorem \ref{Th1}. Cases 8, 9, and 10.} 
\label{connect3}
\end{figure}

The possible configurations are displayed in Fig. \ref{connect3}.
We have $c'=c$ and $h'=h-1$.

In Case 8a we have $b'=b+1$, $v'=v$, and $t'=t-3$ (Fig.  \ref{connect3} (8a)).
In Case 8b we have $b'=b+2$, $v'=v$, and $t'=t-2$ (Fig.  \ref{connect3} (8b)).
In Case 8c we have $b'=b+3$, $v'=v$, and $t'=t-1$ (Fig.  \ref{connect3} (8c)).
In Case 8d we have $b'=b+4$, $v'=v$, and $t'=t$ (Fig.  \ref{connect3} (8d)).

\medskip

{\bf Case 9} \ 
{\em $h$ does not change, while $b$ increases  by 1 and $c$ decreases by 1.}

The only possible configuration is displayed in Fig.  \ref{connect3} (9). 
Note that the pixel marked by 1 is not connected to
the component to which the new pixel $P$ belongs. 
We have $h'=h$, $b'=b+1$, $c'=c-1$, $v'=v$, and $t'=t+1$.

\medskip

{\bf Case 10} \ 
{\em $b$ does not change, 
while $h$ increases by 1 and $c$ decreases by 1 (Case 10a),
or $h$ increases by 1 and $c$ decreases by 2 (Case 10b),
or $h$ increases by 2 and $c$ decreases by 1 (Case 10c).}

The two possible configurations for Case 10a are displayed in Fig.  \ref{connect3} (10a).
For the one on the top we have $v'=v$ and $t'=t+2$, while for the other (bottom) we have $v'=v+1$ and $t'=t+3$.
Case 10b features two possible configurations displayed in Fig.  \ref{connect3} (10b).
For both we have $v'=v$, $t'=t+4$.
In Case 10c we have $v'=v$, $t'=t+4$ (Fig.  \ref{connect3} (10c)).

Note that in all figures numbered pixels belong to a component that is not connected to the pixel $P$.

\smallskip

Having all possible cases determined, simple substitutions in (\ref{eq2}) for the respective values of
$v'$, $t'$, $h'$, $b'$, and $c'$ show that this last equality holds in all cases.

\smallskip

It is easy to realize that the considered cases are the only possible (up to certain symmetries).
Simple reasoning reveals that adding a pixel to $D$ can neither result in
decreasing $b$, nor in
increasing both $h$ and $c$, nor in
increasing both $b$ and $c$, nor in
decreasing $h$ and changing $c$.
This completes the proof of the theorem.
$\Box$
\begin{corollary}
Let $M$ be a tunnel-free digital object. Then $v-2(p+c-h)+b=0$.
\end{corollary}

\subsection{Tunnels in curves}
A digital curve admits various equivalent definitions \cite{BrimkovKlette}.
One of them is the following.
A {\em simple closed digital curve} is a set $\rho = \{c_1,c_2,\dots,c_l \}$ of pixels that satisfy 
the following two axioms: 
{\rm (A1)} $c_i$ is $\alpha$-adjacent to $c_j$  iff  $i = j \pm 1
({\rm modulo} \ l)$, and
{\rm (B1)} $\rho$ is one-dimensional with respect to $\alpha$-adjacency, for $\alpha = 0$ 
(Fig. \ref{connect0}a, top-left) or 1 (Fig. \ref{connect0}a, bottom-left).
To get acquinted with the classical definition of dimension of a digital object the reader 
is referred to \cite{MYL1971}. 
For further developments and various results see \cite{KLE2004,BrimkovKlette} and the
bibliography therein. 
For instance, we have the following fact.
\begin{fact}
\label{F1}
Let a digital object $M$ be one-dimensional with respect to adjacency $\alpha \in \{0,1\}$.
Then $M$ does not contain a 2-block.  
\end{fact}

Any connected subset of a closed digital curve is a {\em simple digital arc} (Fig. \ref{connect0}a, right).
More in general, by analogy to the classical definition of a curve 
in the plane\footnote{A curve in $\R^2$ is one-dimensional continuum, where {\em continuum} is any nonempty subset of a ceratin 
topological space that is compact and topologically connected \cite{Urysohn,Menger}.}, 
a digital curve can be defined as a digital object that is connected and one-dimensional with respect to
an adjacency relation $\alpha$ (see Fig. \ref{connect0}b). 
Applied to digital curves, Theorem \ref{Th1} and Fact \ref{F1} imply the following corollaries.

\begin{corollary}
\begin{itemize}
\item
Let $M$ be a digital curve. Then $t=v-2(p+1-h)$.
\item
If $M$ is tunnel-free, then $v=2(p+1-h)$.
\item
If $M$ is a simple digital arc, then $t=v-2(p+1)$.
\item
If $M$ is a simple tunnel-free digital arc, then $v=2(p+1)$.
\item
If $M$ is a simple closed digital curve, then $t=v-2p$.
\item
If $M$ is a simple closed tunnel-free digital curve, then $v=2p$.
\end{itemize}
\end{corollary}

\section{Final remark}
\label{sec:5}

In this paper we proposed a formula for the number of tunnels in a digital object.
Work in progress is pursuing extension of this result to higher dimensions. 


\end{document}